# Spatial Fluctuations of Helical Dirac Fermions on the Surface of Topological Insulators

Haim Beidenkopf[1], Pedram Roushan[1], Jungpil Seo[1], Lindsay Gorman[1], Ilya Drozdov[1], Yew San Hor[2], R. J. Cava[2], Ali Yazdani[1]

[1]Joseph Henry Laboratory, Department of Physics, Princeton University, Princeton, New Jersey 08544, USA

[2]Department of Chemistry, Princeton University, Princeton, New Jersey 08544, USA

**Surfaces of topological insulators host a new class of states[1] with Dirac dispersion[2-4] and helical spin texture[5]. Potential quantum computing and spintronic applications using these states require manipulation of their electronic properties at the Dirac energy of their band structure by inducing magnetism or superconductivity through doping and proximity effect[6-9]. Yet, the response of these states near the Dirac energy in their band structure to various perturbations has remained unexplored. Here we use spectroscopic mapping with the scanning tunneling microscope to study their response to magnetic and non-magnetic bulk dopants in $Bi_2Te_3$ and $Bi_2Se_3$. Far from the Dirac energy helicity provides remarkable resilience to backscattering even in the presence of ferromagnetism. However, approaching the Dirac point, where the surface states' wavelength diverges bulk doping results in pronounced nanoscale spatial fluctuations of energy, momentum, and helicity. Our results and their connection with similar studies of Dirac electrons in graphene[10-13] demonstrate that while backscattering and localization are absent for Dirac topological surface states, reducing charge defects is required for both**

**tuning the chemical potential to Dirac energy and achieving high electrical mobility for these novel states.**

Since the recent discovery of topological insulators modifications of the bulk chemical compositions or deposition of surface adsorbates have been commonly used to tune their chemical potential into the bulk gap and close to the Dirac energy[3,14-18]. Such efforts are motivated not only by attempts to maximize the ratio of surface to bulk conductivity, but also by the theoretical proposals for realizing yet more exotic states such as Majorana Fermions based on topological insulators[6-9]. Simultaneous proximity to magnetism or superconductivity as well as tuning the chemical potential to the Dirac point of the surface state band structure are required to test these important predictions. It is often assumed that chemical doping that is used to manipulate these states would not disrupt transport of topological surface states because of their topological protection against backscattering. However, experiments on graphene, which is by now a well-studied electronic Dirac system, have demonstrated that nearby defects (adsrobates or those in the substrate) can result in random local gating of such two-dimensional systems. This behavior results in residual concentration of induced carriers in graphene hence preventing the chemical potential from uniformly reaching the Dirac energy throughout the system and scattering that also limits the mobility of Dirac electrons in that system[19,20]. Whether such phenomena are also relevant for topological Dirac surface states that are protected by helical spin texture (as opposed to sub-lattice pseudospin texture in graphene) is not yet known. Finally, it is not clear whether such perturbations, which are non-magnetic and electrostatic in nature, are more dominant than scattering from magnetic defects for the properties of spin helical topological surface state. Addressing these issues is therefore critical for both unraveling the fundamental properties of topological surface state and also finding effective methods for using these states for various applications.

Spectroscopic mapping with the scanning tunneling microscope (STM) has proven a powerful technique to examine the spatial structure and scattering properties of topological surface states in several different topological insulators, from $Bi_{1-x}Sb_x$[21] to those having a single Dirac cone in their surface band structure, such as $Bi_2Se_3$ and $Bi_2Te_3$[22,23]. For non-magnetic scatterers, these experiments have shown evidence for the absence of backscattering between equal and opposite momentum states due to their helical spin texture. Unusual transmission of these states through crystalline barriers has also been reported[24]. Previous STM studies, however, have not probed the properties of topological surface states near the Dirac point of the surface band structure, in part due to the low density of states in the vicinity of that energy[22,23]. The absence of any signatures of scattering in STM measurements near the Dirac energy has been attributed to topological protection of these states against scattering[22,23], although only direct backscattering is strictly prohibited. Whether these states show an unusual sensitivity to magnetic scattering[25] (due to breaking of time-reversal symmetry) is also unsettled as there has not been any STM study contrasting such scatterers to non-magnetic defects.

Our experiments were performed using a home-built cryogenic STM that operates at 4 K in ultrahigh vacuum. Single crystals of $Bi_2Te_3$ and $Bi_2Se_3$ doped with either Ca or Mn were cleaved *in situ* prior to STM measurements taken at low temperatures. Figure 1 shows STM topography and energy-resolved conductance (dI/dV) maps for the different samples at various energies. All topographic images taken on the different compounds (Fig. 1a) show triangular depressions that are associated with the substitutional dopants in layers close to the sample surface[26]. These defects scatter the surface states and give rise to energy-dependent spatial interference patterns displayed in the conductance maps of Fig. 1b and 1c. At low energies (Fig. 1d) the conductance maps are dominated by quenched spatial modulations of the local density of states (LDOS) on a length scale of a few tens of nanometers. Such features are ubiquitous to all doped topological insulators studied here and are present regardless of whether the dopants are magnetic or not.

To determine the microscopic origin of the inhomogeneous LDOS of the topological surface states we show in Fig. 2a and 2b the STM spectra obtained over a range of energies at different locations on the doped $Bi_2Te_3$ and $Bi_2Se_3$ surfaces, respectively. From these spectra it is evident that both the bulk and surface state bands are rigidly shifted in energy between different locations on the surface. We attribute these changes to bulk disorder, such as poorly screened charged deep dopants, as they are not necessarily correlated with the locations of the surface dopants resolved in STM topographs. The Dirac energy in the surface state electronic structure varies spatially together with the bulk-induced nanoscale fluctuations of the electronic band structure, as illustrated by the line in Fig. 2c. The extent of these variations can be mapped by extracting the energy shifts of the spectra obtained at different locations on the sample as shown in Fig. 2d. Our measurements show that the Dirac energy follows a Gaussian distribution with a width of between 20 to 40 meV depending on the specific dopant and its actual concentration. A cross-section of the bulk-induced energy landscape, shown in Fig. 2e, highlights that by tuning the sample's chemical potential close to the Dirac energy, the local electronic structure alternates between electron-like and hole-like doped regions that would have opposing helical spin texture.

The "puddled" real space electronic structure of the Dirac surface states are analogous to similar features observed in graphene by various scanning probe techniques[10-12]. Before describing a more quantitative analysis of these features and their comparison to similar behavior in graphene, we show that these bulk-induced fluctuations have an important influence on the transmission of topological surface states. We determine the impact of bulk defects by examining how the underlying bulk-induced potential affects the response of these states to near-surface defects. Figure 3 shows the short wavelength quasi-particle interference (QPI) patterns for non-magnetically doped $Bi_2Te_3$ and for magnetically doped $Bi_2Te_3$ and $Bi_2Se_3$. Remarkably, detailed Fourier analysis of the QPI patterns shows that short-wavelength scattering on the different samples is independent of whether the scattering dopant is magnetic or not (Fig. 3f-j, lower panels and in Fig. S1) or even if magnetic order is established[26] (see Fig.

S2). In fact, the strong resemblance of the QPI patterns and their discrete Fourier transforms (DFT) across dopants (Ca and Mn) and materials ($Bi_2Te_3$ and $Bi_2Se_3$) suggests that they can be understood based on the shape of the Fermi surface and the spin texture associated with it[2-5,27-29] (Fig. 3k). The warped shape of the surface band structure in $Bi_2Te_3$ and $Bi_2Se_3$ dominates scattering at high energies relative to the Dirac point. Fermi surface warping supports various approximate-nesting wavevectors, but it also induces an out-of-plane component to the spin texture that strongly attenuates the scattering peaks along the $\Gamma$-K direction[29]. Both of these can be accounted for in the calculated spin-dependent scattering probability (SSP)[21,24] presented in Fig. 3l (see Fig. S3 for details). At low energies close to the Dirac point, in the region where the dispersion is conic, the scattering results in a simple circular pattern in the DFT of the real space QPI (Fig. 3i, j). These low energy scattering patterns are also consistent with the helical spin texture of the topological surface states that still allows all scattering processes other than direct backscattering[30], illustrated in Fig. 3m (See also supplementary section). Observation of these features requires improved signal-to-noise ratio when compared to QPI associated with the warped Fermi surface at higher energies.

To determine the influence of bulk-induced disorder potential on the short-wavelength interference patterns caused by the surface impurities, we divide the real-space dI/dV maps into sub-regions, as shown in Fig. 4a, according to intervals of energy-shift of the Dirac point in Fig. 2d. Obtaining the DFT from QPI measurements on two distinct regions (defined by upper and lower halves of the distribution of Dirac energies) we find clear shifts of the QPI's q vectors. An example of such a momentum shift ($\Delta q \sim 0.01$ Å$^{-1}$) is shown in Fig. 4b. Clearly, the surface Dirac electrons alter their wavelength to adjust to the underlying bulk disorder potential, and are not immune to such perturbations. While such fluctuations in momentum are a relatively weak perturbation on the Dirac electrons at high energy, near the Dirac point they are comparable to the average value of the momentum. Figure 4c indeed shows that the Dirac electrons exhibit such a shift in momentum even close to the Dirac energy where their dispersion is perfectly linear. Combining these results with the Fermi velocity obtained from the QPI dispersion ($v_F$=1.3 eV Å, in agreement with ARPES[4]) confirms that the extent of shift in momentum is consistent with the energy shift measured for the Dirac point ($\Delta E = v_F \Delta q \sim 16$ meV). Therefore, at energies close to the Dirac point in the presence of such fluctuations the sample is effectively made up of p-n junctions[31] and momentum carried by these electrons becomes ill defined. The signature of this phenomenon can be seen in the absence of QPI patterns approaching the Dirac point when the topological surface states' wavelength is comparable or larger than the length scale imposed by the bulk-induced disorder. The lack of well-defined momentum near the Dirac point due to the fluctuations reported here is also likely to play a role in the apparent suppression of the ARPES-measured density of states in magnetically-doped topological insulators advocated recently[15-17]. Moreover, the similarity between fluctuations of near-Dirac electrons in both magnetically and non-magnetically doped samples signifies that the bulk-induced fluctuations can have far greater influence (fluctuation variance of 10 meV to be compared with magnetic exchange energy) on the properties of near Dirac electrons than the magnetic nature of scatterers.

The strong fluctuations of the electronic states' momenta near the Dirac point induced by the bulk disorder raises the question of whether such disorder can localize topological surface states[30]. The absence of resonances in the STM spectra (Fig. 2a-c), which one normally expects for electrons confined by random potential on the nanoscale[12,24,32], suggests that the Dirac electrons on topological insulator surfaces are not being localized by the bulk-induced disorder. Considering the typical length scale of the bulk-induced potential, of about 200 Å, we expect electronic states confined on such a length scale to show a level spacing of about 70 meV, which should be well resolved in our STM spectroscopy measurements. The absence of any corresponding resonances in the data implies that the escape rate of the states from puddles is at least comparable to their level spacing, hence imposing a lower bound of 92% for the transmission of topological surface states through the random potential. These results put an upper bound on the possibility of backscattering, in a similar fashion as a previous study on topological surface states in Sb[24]. Such an unusually high value may be indicative of Klein tunneling – a phenomenon unique to Dirac particles in which they fully transmit through an arbitrarily large potential barrier[33,34].

We can carry out a more quantitative analysis of the influence of charge "puddles" caused by bulk disorder in topological surface states, if we apply the theoretical model developed for similar phenomena in graphene[13]. In topological insulators, the bulk defects play the same role as those in the substrate supporting the graphene, to cause fluctuation of the local band structure. Model calculation shows that the variance of the Dirac energy caused by $n_{imp}$ concentration of charged defects at a distance d away from the sample surface can be computed using the relation,

$$\Delta E_D = \sqrt{2\pi n_{imp} C_o}\, \frac{e^2}{\kappa},$$

where $\kappa$ is the bulk dielectric constant, and $C_o(r_s, k_F d)$ depends on surface Dirac systems' carrier concentration characterized by $r_s$ (0.24) and the Fermi wavelength $k_F$ (0.1 Å$^{-1}$). Assuming the relevant defects lie within d~10 Å from the surface we find[13] $C_o$~0.5. Since the bulk dopants are randomly distributed, we can reasonably estimate their density by counting the number of substitutional defects in the upper most Bi layer (2.5 Å below the surface) from STM topographic images ($n_{imp}$~4x10$^{12}$ cm$^{-2}$). Finally, due to screening geometry (half plane) and the heavy doping of our samples, we use reduced value of the bulk dielectric constant of 30 (compared to 113 in $Bi_2Se_3$ and 75-290 in $Bi_2Te_3$[35]). With these parameters, we arrive at an estimate of 16 meV for the variance of the Dirac point, which agrees well with our experimental results based on analysis of STM data (Fig. 2a). The puddles' spatial correlation length can also be estimated by the Thomas-Fermi screening length scale in a Dirac system $L_{TF} = 2\pi\kappa\hbar v_F / e^2 k_F$ to be about 260 Å, which is also in good agreement with that extracted from our spectroscopic maps of the Dirac point in Fig. 2d.

The physical picture emerging from our experiments and their analysis is that despite their protection against backscattering and localization, topological surface states are susceptible to charge disorder caused by bulk defects. This conclusion is consistent with recent transport studies on topological insulators claiming that doped samples, besides having low mobilities, show magneto-fingerprints associated with mesoscopic fluctuations on the length scale of a few tens of nanometer found in macroscopic single crystals[18]. Transport experiments using electrostatic gating on thin films give evidence of a minimum conductivity that has been attributed to a defect induced residual charge density[36]. Both studies are consistent with our observation of bulk-induced fluctuations that introduce a length scale associated with screening of the defects and prevent from tuning the chemical potential to the Dirac energy because of defect-induced charge puddles. Overall, similar to efforts in graphene, reduction of bulk charge defects appears to be required to fully realize the potential of topological insulators to carry out fundamental studies associated with properties of this system near the Dirac point and for any potential applications that relies on high carrier mobility. Our quantitative analysis of puddles' size and amplitude here suggest that the large bulk dielectric constant of topological insulator provide these systems with a natural advantage. In sufficiently clean samples, we expect the fluctuation in the electronic structure to be ultimately much weaker as compared to those reported here.

**Acknowledgements**

We gratefully acknowledge discussions with R. R. Biswas, S. Adam, and N. P. Ong. This work was supported by DARPA, ONR, NSF-DMR, and through NSF-MRSEC program through Princeton Centre for Complex Materials. The instrumentation and infrastructure at the Princeton Nanoscale Microscopy Laboratory are also supported by grants from DOE, the W.M. Keck foundation, and Eric and Linda Schmidt Transformative fund at Princeton.

**Figure captions:**

**Figure 1. Interference and inhomogeneity on the surface of doped topological insulators. (a)** Topographic images on the surface of Mn- and Ca-doped $Bi_2Te_3$ and Mn-doped $Bi_2Se_3$. **(b-d)** Conductance (dI/dV) maps taken at high, intermediate and low sample biases showing dispersive QPI patterns that ride a quenched inhomogenity of the LDOS.

**Figure 2. Bulk origin of the charge inhomogeneity detected on the surface. (a,b)** The LDOS measured at various locations on the surface of Mn-doped $Bi_2Te_3$ and $Bi_2Se_3$, respectively, showing a rigid shift of the bulk and surface bands (band structures shown schematically). The Dirac point, $E_D$ (approximated by linear extrapolation of the surface state's LDOS in $Bi_2Te_3$, and as the point of minimal LDOS in $Bi_2Se_3$), shifts with them. Insets show the Gaussian distribution of $E_D$. **(c)** High energy resolution LDOS taken along a line cut that crosses several charge puddles (white line) on Mn-doped $Bi_2Se_3$ demonstrating absence

of resonances. **(d)** Spatial distribution of $E_D$ extracted from local shift in LDOS. Its structure shows no correlation with the locations of dopants (triangles). **(e)** Line cut along yellow line in (d) showing electron and hole "puddles" that would form once the chemical potential is tuned to the average Dirac point in the sample.

**Figure 3. Scattering of surface states. (a-e)** Real-space QPI patterns on the surface of Ca-doped $Bi_2Te_3$ at different energies. **(f-j)** Fourier transforms of the QPI patterns from Ca- and Mn-doped $Bi_2Te_3$ and Mn-doped $Bi_2Se_3$ given in top, middle and bottom panels, respectively. All compounds show similar patterns in q-space consisting of six strong peaks along the $\Gamma$-M directions at high energies and circular patterns at lower ones. **(k)** Schematic surface band-structure in $Bi_2Te_3$ and the associated spin texture. Warping at high energies supports approximate-nesting conditions. Those along $\Gamma$-M are indicated by green arrows. **(l)** Calculated spin dependent scattering probability (SSP) captures the Fourier space QPI patterns. **(m)** Illustration of processes and their contribution to the QPI pattern: helicity forbids backscattering off non-magnetic impurities (top) but allows oblique scattering and interference that have a finite overlap between initial and final spin states (middle). A magnetic impurity allows spin-flip backscattering but not interference of initial and final spin states that remain orthogonal.

**Figure 4. Spatial fluctuations of momentum. (a)** The conductance map on $Bi_{1.95}Mn_{0.05}Te_3$ divided into sub-regions with high and low LDOS. **(b)** Profile of a q-space QPI peak along $\Gamma$-M (gray), and those peaks when Fourier transformed separately from the different sub-regions in (a) showing a relative shift in momentum $\Delta q=0.01$ Å$^{-1}$ (lines are Gaussian fits to the data points). **(c)** Similar

analysis in the region of conical energy dispersion of Bi$_{1.95}$Mn$_{0.05}$Se$_3$. QPI peaks

from both sub-regions show linear dispersion shifted by energy independent

$\Delta q \sim 0.01$ Å$^{-1}$ with respect to each other. The dashed line marks $q=2\pi/L$ below

which the electronic wavelength is greater than the typical dimension of a puddle,

$l$, and the QPI patterns vanish.

**Figures**

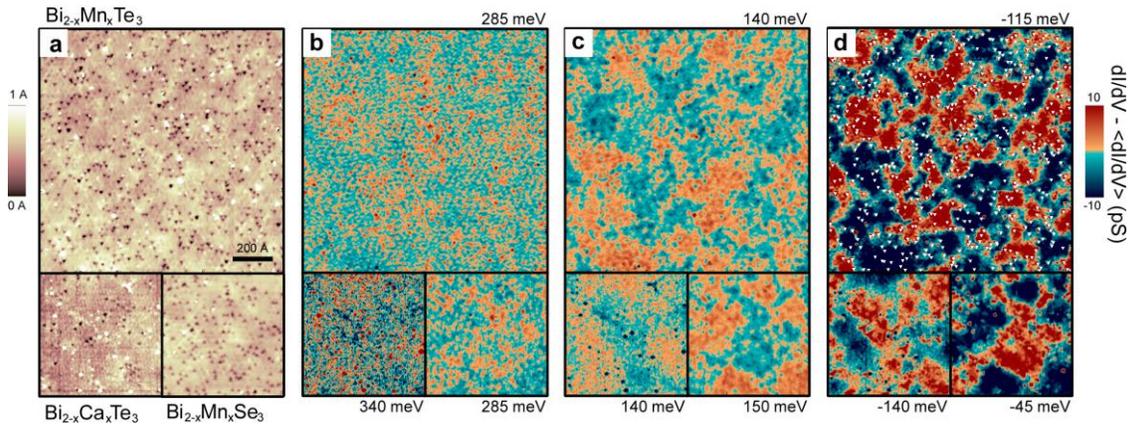

Figure 1

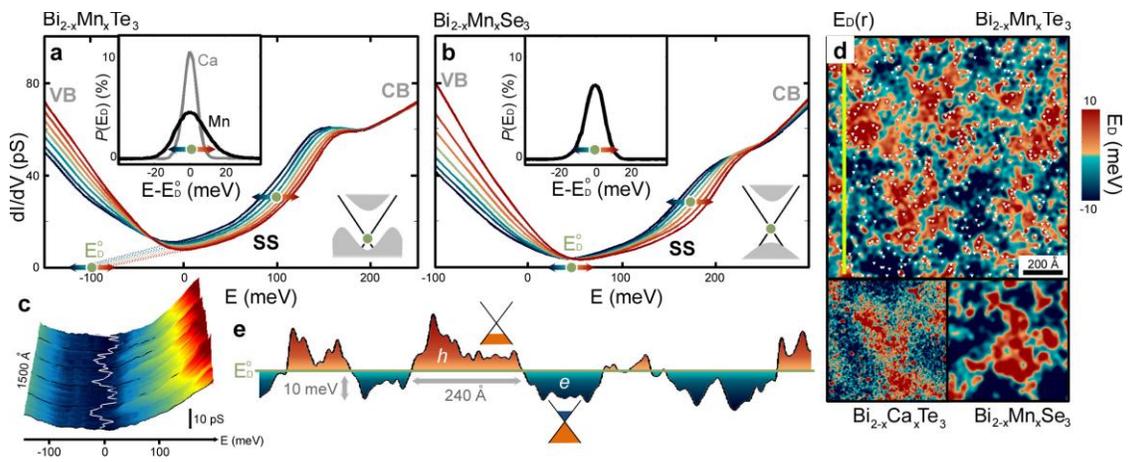

Figure 2

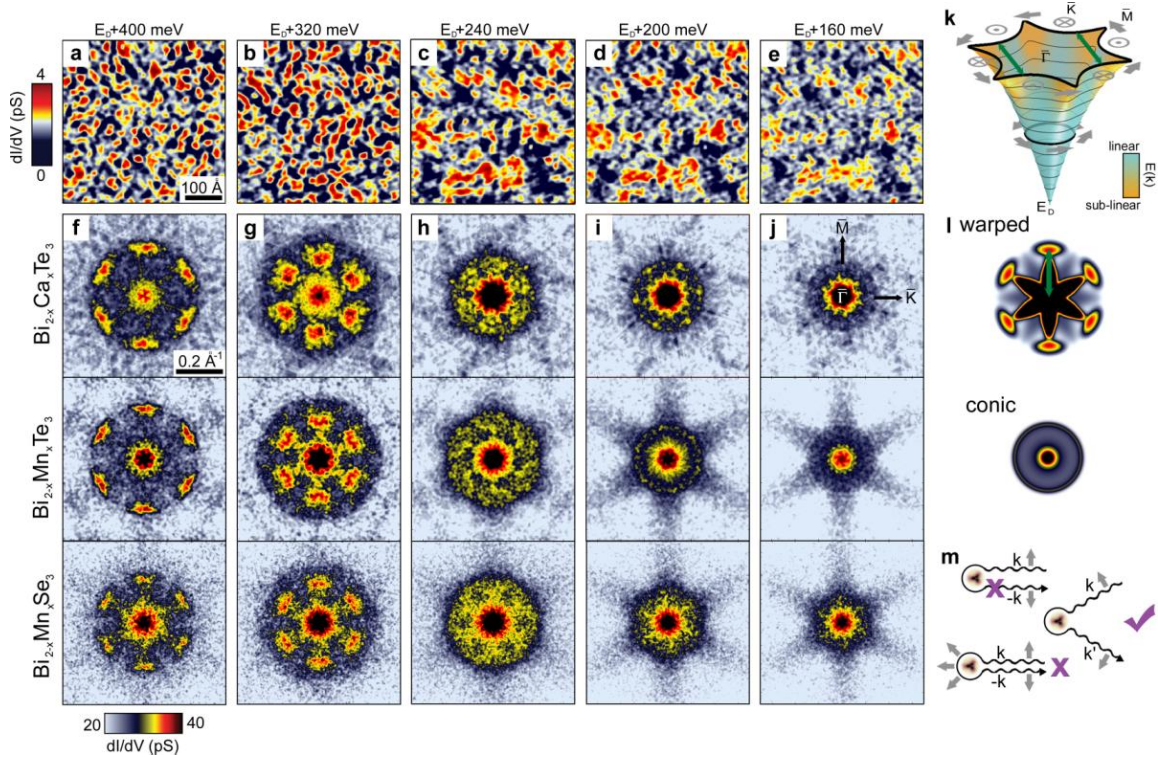

Figure 3

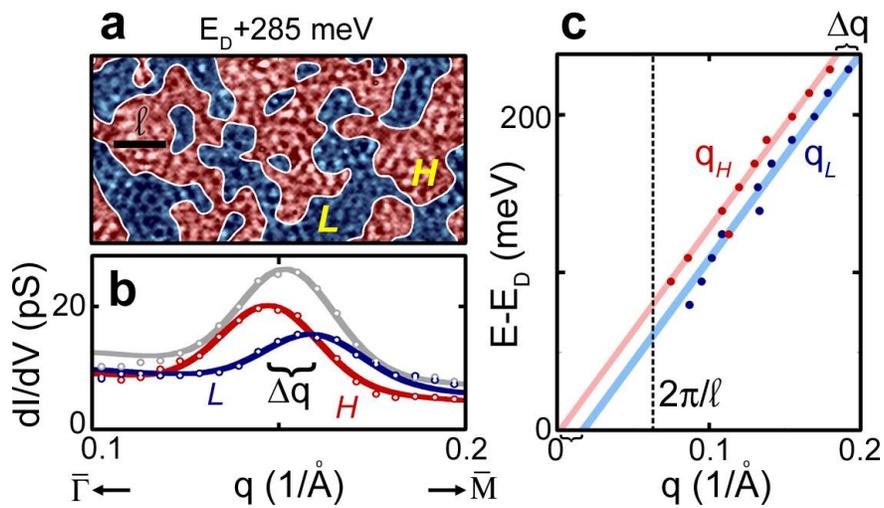

Figure 4

**Supplementary Information**

**1. Movies of dispersing QPI patterns and their DFTs on $Bi_{1.95}Mn_{0.05}Te_3$ and $Bi_{1.95}Mn_{0.05}Se_3$.**

**2. Scattering and magnetism**

Since the partial protection from scattering is assured by time reversal symmetry, magnetism is believed to have a dominant impact on the allowed scattering processes of the helical surface states. Once the impurity acquires a magnetic moment it can spin-exchange with the incident Dirac particle, and allow for spin-flip backscattering. Nevertheless, the QPI patterns on the magnetically doped compounds, $Bi_{1.95}Mn_{0.05}Te_3$ and $Bi_{1.95}Mn_{0.05}Se_3$, given in the central and lower panels of Figure 3(f-j), respectively, show striking similarity to those we find in the non-magnetic Ca-doped $Bi_2Te_3$ shown in the upper panels of the figure. Even the $\Gamma$-K scattering wavevctors at high energies, that get suppressed by the spin texture, remain absent in the magnetically-doped compounds. We note that although spin-flip backscattering should be facilitated in Mn-doped samples, in order for STM to be able to detect backscattering events they must also alter the LDOS through constructive interference. As long as the helical spin texture remains intact the orthogonality of the spin states prevents the backscattered states from doing so, as illustrated in the bottom panel of Figure 3(m). Therefore, it is the remaining continuum of possible oblique scattering channels, whose spin states are not fully orthogonal, that gives rise to the measured interference patterns, as is the case in the non-magnetic Ca-doped $Bi_2Te_3$.

One may further speculate that breaking time reversal symmetry locally at Mn sites, or even more so globally within a ferromagnetic phase that Mn-doped $Bi_2Te_3$ indeed exhibits below $T_C=10$ K, can distort the spin texture and allow all channels to interfere. However, the low ferromagnetic transition temperature indicates a gap on the order of a few meV which opens at $E_D$. Such a small gap can degrade the spin texture only on an energy scale of its order. In contrast, spin-orbit coupling in these materials, which brings about the helical spin texture, has a much larger energy scale of several eV. Therefore, magnetism is too weak to account for the interference patterns we detect throughout the ~250 meV that span the conic region of the dispersion. In accordance, the QPI patterns measured at 2 K, far below $T_c$, are essentially identical to those taken above it at 20 K as movie S3 clearly shows.

## 3. Origins of the QPI patterns.

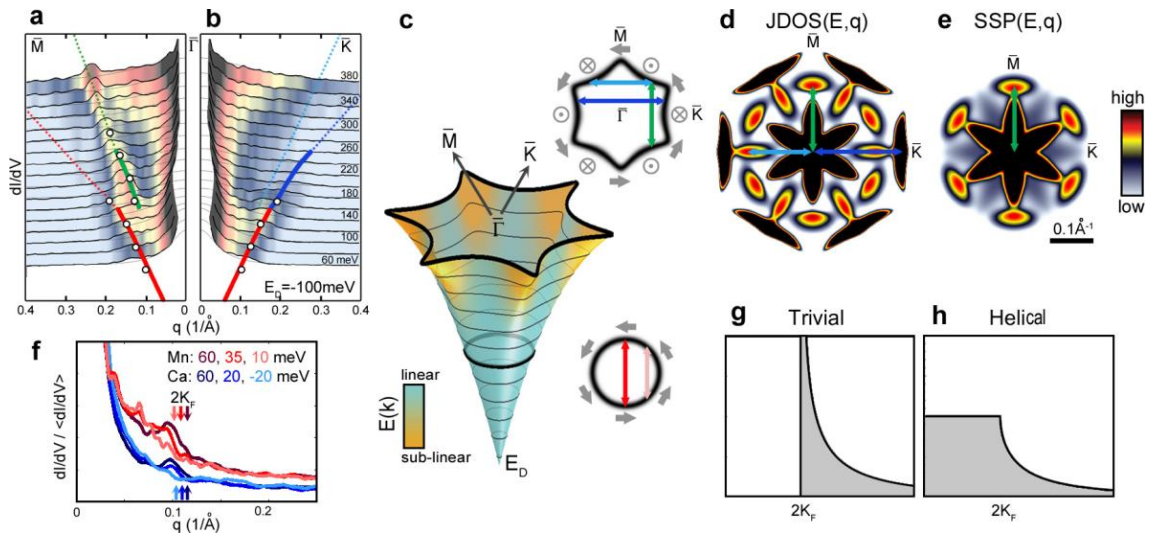

**Figure S1. QPI patterns based on spin texture.** The Full dispersion of the scattering wavevectors in **(a)** $\Gamma$-M, and in **(b)** $\Gamma$-K direction, obtained from the Fourier transform of the dI/dV maps taken on Mn-doped $Bi_2Te_3$ are shown in color. They agree with those measured on Ca-doped $Bi_2Te_3$ (circles) as well as with the dispersing q-vectors calculated based on ARPES (red, green and blue solid lines, and their linear extrapolation in dotted lines). **(c)** The band structure of the surface states in $Bi_2Te_3$, as measured by ARPES, turns at ~$E_D$+140 meV from conic to hexagonally warped. The associated spin texture is given with gray symbols. Scattering peaks (marked with arrows) found in the calculated JDOS **(d)** and SSP **(e)**. **(f)** Azimuthally averaged radial profiles of the QPI patterns in the conic region of the dispersion in Mn- and Ca-doped $Bi_2Te_3$, normalized by the average LDOS at the respective energy. To be compared with the theoretical prediction for asymptotic behavior in a trivial **(g)** and helical **(h)** metals.

In order to trace the origin of the various intensity peaks in the QPI patterns of Fig.3(f-j) we first plot in Fig. S1(a) and (b) cuts taken along the $\Gamma$-M and $\Gamma$-K directions, respectively, out of the full QPI pattern in Mn-doped $Bi_2Te_3$. The isotropic QPI patterns we find at low energies translate to symmetric modes in $\Gamma$-M and $\Gamma$-K. They disperse linearly, and their extrapolation sets the Dirac point at ~100 meV below the Fermi energy. In contrast, the scattering mode we detect at high energies along $\Gamma$-M is completely absent in the $\Gamma$-K direction.

The anisotropic QPI patterns at high energies stem from the anisotropic surface state's band structure, shown schematically in Fig. S1(c). ARPES measurements imply that at energies beyond ~140 meV above the Dirac point the $\Gamma$-M dispersion becomes sublinear. As a result the band structure attains a snow-flake profile that can accommodate various approximate-nested scattering wavevectors both along the $\Gamma$-M and the $\Gamma$-K directions, indicated in Fig. S1(c) with arrows. To identify the dominant scattering wavevctors we calculate the joint density of states JDOS(E,q)=$\int_{BZ}\rho$(E,k)$\rho$(E,k+q)dk$^2$ with $\rho$(E,k) being the energy and momentum resolved LDOS. Strong peaks along both $\Gamma$-M and $\Gamma$-K indeed appear in the calculated JDOS shown in Fig. S1(d). The dispersion of these dominant scattering wavevectors with energy, based on ARPES measured dispersion of $\rho$(E,k), is given by solid lines in Fig. S1(a,b) (the dotted lines being their linear extrapolation to energies ARPES cannot access).

We attribute the absence of the strong $\Gamma$-K scattering peaks from the measured QPI patterns mainly to the protection cast at these high energies by the full spin texture. While spin-orbit generates the in-plane spin texture, the hexagonal warping term has to couple to the out-of-plain spin component. This cants the spin out of plane in an alternating fashion only along $\Gamma$-K demonstrated by the gray symbols in panel (d). The out-of-plane spin component provides additional protection against backscattering in the $\Gamma$-K direction alone. This can be captured by calculating the spin-dependent scattering probability SSP(E,q)=$\int_{BZ}\rho$(E,k)$|<\sigma_K|\sigma_{k+q}>|^2\rho$(E,k+q)dk$^2$ where $|\sigma_k>$ denotes the helical spin state. In the resulting SSP, shown in Fig. S1(e), the $\Gamma$-K nesting peaks are

absent, and only those along Γ-M remain. The calculated SSP looks identical to the QPI pattern we find at high energies [see Fig.3 (f) for instance].

Naturally, the out-of-plane spin component vanishes at low energies together with warping, rendering the spin-selection rules isotropic in accordance with the circularly symmetric QPI pattern we encounter at low energies. The low-energy QPI radial profile is plotted in Fig. S1(f) for Mn- and Ca-doped $Bi_2Te_3$. In both cases a $2k_F$ feature appears on top of a broad central peak which results from the inhomogeneous charge-puddle structure in real-space. The feature at $2k_F$ disperses inwards with decreasing energy in a rate that corresponds well with $v_F$. Theoretically, the exact shape of the $2k_F$ feature is affected by the presence or absence of helical spin texture. The standing wave pattern that forms next to a point-impurity in a metal with circular Fermi surface decays asymptotically as $cos(2k_F r)/r$, whose DFT has a divergent $2k_F$ peak[37] [Fig. S1(g)]. In contrast, the standing wave pattern in a helical metal decays faster, as $sin(2k_F r)/r^2$, whose DFT shows a $2k_F$ cusp rather than peak[30,38] [Fig. S1(h)]. It should be noted, though, that a transition from helical to trivial QPI profile occurs only under strong ferromagnetism that induces a sizable gap at the Dirac energy[38]. Weakly correlated magnetic impurities should not alter the QPI pattern measured by (non-spin polarized) STM as they can spin-flip backscatter but cannot give rise to constructive interference. The feature we find in experiment is far weaker than that reported in the case of Shockley states on metal surfaces[39]. Riding the strong background further limits our ability to draw a distinction between a peak and a cusp, though it does seem like both in Mn- and Ca-doped cases a faint $2k_F$ peak appears at 60 meV, and decays to a cusp at lower energies. A faint peak may be attributed to a residual weak warping that would result in a non-zero out-of-plane spin component, or to a partial spin protection due degraded or weakly fluctuating spin texture.

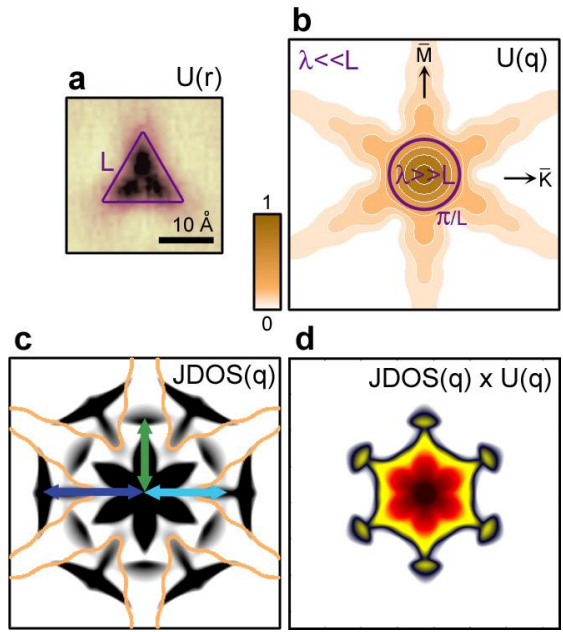

**Figure S2. QPI patterns based on scatterer geometry. (a)** Topographic image of a single Mn dopant in Bi2Te3. **(b)** Fourier transform of a triangle of dimension L. **(c)** Warped JDOS superimposed by the short-wavelength anisotropic pattern of (b). **(d)** The warped JDOS once modified by the geometry of the scattering potential.

We further note that the shape of the dopants once embedded in the lattice and their dimension as captured in the topographic image of Figure S2(a) may also contribute to the suppression of the $\Gamma$-K scattering peaks at high energies. The shape of the scatterer enters the JDOS as a prefactor, $JDOS(E,q)=V(q)\int_{BZ}\rho(E,k)\rho(E,k+q)dk^2$, where $V(q)$ is the Fourier transform of the scattering potential $V(r)$ off a single dopant. Approximating $V(r)$ as a triangle of dimension L, demonstrated in Figure S2(a), yields the Fourier-space pattern for $V(q)$ shown in Figure S2(b). While the long wavelength scattering ($\lambda \gg L$) is insensitive to the details of the scatterer, the short wavelength scattering processes ($\lambda << L$) are greatly affected by the scatterer anisotropy. In particular, the $\Gamma$-K scattering gets suppressed relative to the $\Gamma$-M direction. By superimposing the short wavelength (large q) anisotropic pattern of V(q) over the warped JDOS, shown in Figure S2(c), we find that it is exactly the $\Gamma$-K scattering peaks of the JDOS that would be attenuated by the scatterer's triangular shape. This is indeed the resulting JDOS once modified by the approximated V(q) term, as demonstrated in Figure S2(d). Finally, the crossover from isotropic-like long wavelength pattern to the anisotropic long wavelength one will occur at $q \sim \pi/L = 0.2$ Å$^{-1}$ for L~15 Å. Such a wavevector would suppress $\Gamma$-K scattering in the warped region of the band structure while leaving the scattering in the isotropic conic region unaltered in agreement with what we find in QPI. We thus find that both the spin-texture as well as the geometry of the dopants contribute to the suppression of the measured $\Gamma$-K peaks in the Fourier-space QPI patterns.